\documentstyle[prl,aps,hangcaption,psfig]{revtex}
\begin{document}
\begin{twocolumn}
\wideabs{

\title{{\small \rightline{NUC-MINN-01/8-T}}
Second Order Dissipative Fluid Dynamics
for Ultra-Relativistic Nuclear Collisions}
\author{Azwinndini Muronga\cite{byline}}
\address{ School of Physics and Astronomy, University of Minnesota, 
         Minneapolis, Minnesota 55455, USA}
\date{\today}
\maketitle

\begin{abstract}

The M\"uller-Israel-Stewart second order theory of
relativistic imperfect fluids based on Grad's moment method is used to study the
expansion of hot matter produced in ultra-relativistic
heavy ion collisions. The temperature evolution is investigated in the framework
of the 
Bjorken boost-invariant scaling limit. The results of these second-order theories
are
compared to those of first-order theories due to Eckart and to Landau and Lifshitz
and those of zeroth order (perfect fluid) due to Euler.
\end{abstract}
}
\vspace{0.3cm}
High energy heavy-ion collisions offer the opportunity to study the 
properties of hot and dense matter. To do so we must follow its 
space-time evolution, which is affected not only by the equation of state but
also by dissipative, non-equilibrium processes. Thus
we need to know the transport coefficients
such as viscosity, thermal conductivity, and diffusion. We also need to
know the relaxation coefficients. Knowledge of the various time and length scales is of central
importance to help decide 
whether to apply fluid dynamics or cascade or a combination
of the two.
The use of fluid dynamics as one of the approaches in modeling the dynamic
evolution of nuclear collisions has been successful in describing many of the 
observables \cite{azwi:stocker1986,azwi:kajantie1986}.  So far most work has
focused on the ideal or perfect fluid and/or
multi-fluid dynamics.
In this work we apply the relativistic dissipative fluid dynamical approach. 
It is known even from non-relativistic studies \cite{azwi:kapusta}
that dissipation might affect the observables.
The first
theories of relativistic dissipative fluid dynamics are due to Eckart \cite{azwi:eckart1940}
and to Landau and Lifshitz \cite{azwi:landau1959}. These theories are now known to
have some undesirable features: they lead to Navier-Stokes-Fourier laws which are
parabolic in structure and therefore may propagate signals with
speeds exceeding that of light. 
A qualitative study of relativistic dissipative fluids for applications to 
relativistic
heavy ions collisions has been done using these first-order theories
\cite{azwi:naviergang}.

Second-order theories of dissipative fluids due to Grad\cite{azwi:grad1949},
M\"uller\cite{azwi:muller1967}, and
Israel and Stewart \cite{azwi:israelstewart} were introduced to remedy some
of these undesirable features. In second-order theories the space of thermodynamic
quantities is expanded to include the dissipative quantities for the particular
system under consideration. These dissipative quantities are treated as thermodynamic variables in
their own right.
The phenomenological formulation of the transport equations for the first-order and second-order theories
is accomplished by combining the conservation of
energy-momentum and particle number with the Gibbs equation. One
then obtains an expression for the entropy 4-current, and its divergence leads to
entropy production. 
Because of the enlargement of the space of variables the expressions for the 
energy-momentum tensor $T^{\mu\nu}$, particle 4-current $N^\mu$, entropy
4-current $S^\mu$, and the Gibbs
equation contain extra terms.
Transport equations for dissipative fluxes are obtained by
imposing the second law of thermodynamics, that is, the principle of nondecreasing 
entropy. The difference between the two stems from 
the entropy 4-current: the standard irreversible thermodynamics of
Eckart-Landau assumes that the entropy 4-current should include 
terms linear in dissipative fluxes and
hence they are referred to as {\em first-order theories} of dissipative fluids.
On the other hand the extended irreversible thermodynamics of
M\"uller-Israel-Stewart includes terms quadratic in
dissipative fluxes and hence they are referred to as {\em second-order theories}
of dissipative fluids. The kinetic approach is based on Grad's
14-moment method \cite{azwi:grad1949}. 
The resulting equations are hyperbolic and lead to causal
propagations \cite{azwi:israelstewart,azwi:hiscock}. 
For a review on generalization of the M\"uller-Israel-Stewart theory to a mixture of
several particle species see \cite{azwi:prakash}.

The formulation of relativistic hydrodynamics can be found in standard text books
\cite{azwi:landau1959,azwi:weinberg,azwi:laszlobook,azwi:degroot,azwi:rischke}. 
The energy equation governing fluid motion is given by
\begin{eqnarray}
 D\varepsilon &=& - (\varepsilon+p+\Pi)\theta +\pi^{\mu\nu}\bigtriangledown_\nu u_\mu -\bigtriangledown_\mu q^\mu
 +q^u a_\mu  \enspace, \label{azwi:energy}
\end{eqnarray}
where  
\begin{eqnarray}
\varepsilon \,&&\,\;\; \mbox{(energy density)} \enspace,\nonumber\\
p  \,&&\,\;\; \mbox{(pressure)} \enspace,\nonumber\\
D &\equiv& u^\mu \partial_\mu \,\,\ \,\,\,(\mbox{convective time
derivative}) \nonumber \enspace,\\
\bigtriangledown^\mu &\equiv& \bigtriangleup^{\mu\nu}\partial_\nu \,\,\,\,( \mbox{
gradient operator}) \nonumber \enspace\\
\theta  & \equiv& \partial_\alpha u^\alpha \;\; \mbox{(volume expansion )}
\enspace,\\
u^\mu \,&&\, \mbox{(4-velocity)} \nonumber\enspace,\\
a^\mu\ &\equiv& D u^\mu \,\mbox{(4-acceleration)} \nonumber\enspace,\\
\Pi\,&&\,\;\; \mbox{(viscous pressure)}\nonumber \enspace,\\
\pi^{\mu\nu}\,&&\,\;\; \mbox{(viscous stress tensor)}
\nonumber \enspace,\\
q^\mu \,&&\,\;\; \mbox{(heat flow)}
\nonumber \enspace,
\end{eqnarray}
$\bigtriangleup^{\mu\nu} \equiv g^{\mu\nu}-u^\mu u^\nu$ is the projection tensor
orthogonal to the 4-velocity and $g^{\mu\nu}$ = diag$(+1,-1,-1,-1)$ is the metric tensor
in Minkowski space-time. 
In the Eckart theory the dissipative contribution to the bulk pressure, the heat
flux and the shear viscous
tensor are given by
\begin{eqnarray}
\Pi &=& -\zeta \theta \enspace,\\
q^\mu &=& \lambda T\left(\frac{\bigtriangledown^\mu T}{T}- a^\mu\right) \enspace,\\
\pi^{\mu\nu} &=& 2\eta \bigtriangledown^{\langle\mu}u^{\nu \rangle} \enspace,
\end{eqnarray}
where $\zeta\,,\eta\,,\lambda$ are the bulk, shear and thermal conductivity
coefficients respectively. They are required to be positive by the second law of
thermodynamics. $T$ is the temperature of the system. 
The angular bracket notation is defined by:
\begin{eqnarray}
A^{<\mu\nu>} &\equiv& \left[\frac{1}{2}\left(\bigtriangleup^\mu_\sigma
\bigtriangleup^\nu_\tau
+\bigtriangleup^\mu_\tau \bigtriangleup^\nu_\sigma\right)
-\frac{1}{3}\bigtriangleup^{\mu\nu}\bigtriangleup_{\sigma\tau}\right]
A^{\sigma\tau} \enspace.
\end{eqnarray}
The above expressions for dissipative fluxes can also be obtained in kinetic
theory by employing the first Chapman-Enskog approximation
\cite{azwi:chapman}. 
In second-order theories we have to solve for these dissipative fluxes
from their evolution equations.
One can show \cite{azwi:project} that in the 1+1 Bjorken scaling solution
hypothesis \cite{azwi:bjorken}, the system of transport equations given by 
 \cite{azwi:israelstewart} becomes tractable. The same simplifications occur
in 1+1 dimensional Bjorken hypothesis in first order dissipative fluid
calculations \cite{azwi:naviergang}. In 2+1 and 3+1 dimensional flow these
equations become much more complicated. 
This is a subject of current study \cite{azwi:project}.

I give here the second-order equations for 1+1
dimensional scaling solution. 
These equations were derived from kinetic
theory using Grad's 14-moment approximation method \cite{azwi:grad1949}.
We will only need the M\"uller-Israel-Stewart \cite{azwi:israelstewart} 
equations in the following form:
\begin{eqnarray}
D\Pi &=& -\frac{1}{\tau_\Pi}\Pi 
-\frac{1}{2}\frac{1}{\beta_0} \Pi
\left(\beta_0\theta+TD\left(\frac{\beta_0}{T}\right)\right) \nonumber \\
&&-\frac{1}{\beta_0}\theta \enspace, \label{azwi:bulkeqn}\\
Dq^\mu &=& -\frac{1}{\tau_q}q^\mu
+\frac{1}{2}\frac{1}{\beta_1} q^\mu
\left(\beta_1\theta+TD\left(\frac{\beta_1}{T}\right)\right) \nonumber \\
&&+\frac{1}{\beta_1}
\left(\frac{\bigtriangledown^\mu T}{T}- a^\mu\right) \enspace, \label{azwi:heateqn}\\
D\pi^{\mu\nu} &=& -\frac{1}{\tau_\pi}\pi^{\mu\nu} - \frac{1}{2}\frac{1}{\beta_2}
\pi^{\mu\nu}\left(\beta_2\theta + T D\left(\frac{\beta_2}{T}\right)\right)\\
\nonumber
&&+\frac{1}{\beta_2} \bigtriangledown^{\langle\mu}u^{\nu\rangle}
  \enspace, \label{azwi:sheareqn}
\end{eqnarray}
where 
\begin{equation}
\tau_\Pi=\zeta \beta_0\,\,\,\,,\tau_q = \lambda T \beta_1\,\,\,\,, 
\tau_\pi = 2 \eta \beta_2 \nonumber \enspace,
\end{equation}
are the relaxation times, $\beta_0\,, \beta_1$ and $\beta_2$ are the relaxation
coefficients. These three
new coefficients are functions of primary thermodynamic variables like pressure,
number density and energy density and hence depend on the equation of
state. Relaxation time is the distinguishing feature of
hyperbolic causal theories which is not present in the first-order theories. 
Here $\tau_i$ 
is the time taken by the 
corresponding dissipative flux to relax to its steady-state value.

For the 1+1 dimensional Bjorken \cite{azwi:bjorken} similarity fluid flow the energy equation becomes:
\begin{equation}
\frac{d\varepsilon}{d\tau}+\frac{\varepsilon+p}{\tau}
-\frac{2}{3}\frac{1}{\tau} \Phi +\Pi \frac{1}{\tau} = 0 \enspace,
\label{azwi:scalingenergy}
\end{equation} 
where $\Phi \equiv \pi^{00}-\pi^{zz}$ is determined from the shear viscous tensor evolution equation
\begin{eqnarray}
\frac{d}{d\tau}\Phi &=&
-\frac{1}{\tau_\pi}\Phi - \frac{1}{2}\Phi\left(\frac{1}{\tau} +\frac{1}{\beta_2}
T \frac{d}{d\tau}\left(\frac{\beta_2}{T}\right)\right)
+\frac{2}{3}\frac{1}{\beta_2}\frac{1}{\tau} \enspace. \label{azwi:scalingshear}
\end{eqnarray}
The equation for $\Pi$ will not needed as explained below.
For this initial study a simple equation of state is used, namely that of a weakly interacting plasma 
of massless $u\,,d\,,s\,$ quarks and gluons. The pressure is given by $p =
\varepsilon/3 = aT^4$ with zero baryon chemical potential. Here $a$ is a
constant determined by the number of quark flavors and the number of gluon
colors. In the case of massless particles the bulk pressure
equation (\ref{azwi:bulkeqn}) does not contribute since the bulk viscosity is
negligible or vanishes
\cite{azwi:weinberg}.  
The only relaxation coefficient we need is $\beta_2$ which for massless
particles is given by $\beta_2 = 3/(4 p)$.
The shear viscosity is given by
\cite{azwi:baym} $\eta = b T^3$ where $b$ is a constant determined by the number of
quark flavors and the number of gluon colors. The energy equation (\ref{azwi:energy}) and the viscous stress
tensor equation (\ref{azwi:sheareqn}) can be written as 
\begin{eqnarray}
\frac{d}{d\tau}T &=& -\frac{T}{3\,\tau} + \frac{T^{-3}\Phi}{18\,a\,\tau}
\enspace,\label{azwi:temp}\\
\frac{d}{d\tau} \Phi &=& - \frac{2\,a\,T\Phi}{3\,b} -
\frac{1}{2}\left(\frac{1}{\tau}-5 \frac{1}{T}\frac{d}{d\tau}T\right)\Phi +\frac{8\,a\,T^4}{9\,\tau}
\enspace. \label{azwi:shear}
\end{eqnarray}
where
\begin{eqnarray}
a &=& \left(16 + \frac{21}{2} N_f\right)\frac{\pi^2}{90}\enspace,\\
b &=& \left(1 + 1.70 N_f\right)\frac{0.342}{(1+N_f/6)\,\alpha_s^2\ln(\alpha_s^{-1})}
\enspace,
\end{eqnarray}
Here $N_f$ is the number of quark flavors, taken to be $3$, and
$\alpha_s$ is the strong fine structure constant taken to be $0.5$.
For a perfect fluid and a first-order theory the energy equation
(\ref{azwi:scalingenergy})
can be solved analytically to give
\begin{eqnarray}
T(\tau) &=& T_0\left[\frac{\tau_0}{\tau}\right]^{1/3}
\label{azwi:perfect} ~~~~~\mbox{(perfect fluid)}\enspace,\\
T(\tau) &=& T_0\left[\frac{\tau_0}{\tau}\right]^{1/3} \left\{1+\frac{b}{6\, a}
\frac{1}{T_0\,\tau_0}
\left(1-\left[\frac{\tau_0}{\tau}\right]^{2/3}\right)\right\}
\label{azwi:1st}\\
&&\mbox{(first-order theory)} \enspace, \nonumber
\end{eqnarray}
In the first order theory we do not have the relaxation coefficients. Then
 equation (\ref{azwi:scalingshear}) gives $\Phi = (4 \eta/3)/\tau$.
The above equations and the numerical solution to the second-order equations
(\ref{azwi:temp}) and (\ref{azwi:shear}) are 
presented in Figs. \ref{azwi:Fig1} through \ref{azwi:Fig4}.
We choose the initial temperatures to
correspond to those expected at RHIC and LHC, namely 
$T_0$=500 MeV at RHIC and $T_0$=1000 MeV at LHC. In  Figs.
\ref{azwi:Fig1} and \ref{azwi:Fig2} the
initial time $\tau_0$ is estimated by using the uncertainty principle \cite{azwi:kms}: 
$\tau_0 \cdot {\langle E \rangle}_0 \sim 1$ where ${\langle E \rangle}_0 \sim 3
T_0$ for massless particles. This results in 
$\tau_0$=0.13 fm/c at RHIC and $\tau_0$=0.07 fm/c at LHC. 
The initial value used for $\Phi$, which must be specified independently for
the second order theory, is taken to be $p/3$. We choose this value
since the second order theory is based on the assumption that the dissipative
fluxes are small compared to the primary thermodynamic variables,
namely $p,n, $and $\varepsilon$. However, a thorough study of the initial
conditions on these dissipative fluxes is needed and should perhaps be found
from microscopic models. This is as subject of current study
\cite{azwi:project}.
The effect of
dissipation is more pronounced at the very early stages of heavy ion
collisions when gradients of temperature, 
velocity, etc., are large. 
At late times the effect of dissipation vanishes. In Figs. \ref{azwi:Fig3} and \ref{azwi:Fig4} 
we take a constant initial time 
$\tau_0$=1.0 fm/c which is the characteristic hadronic time scale. 
Euler hydrodynamics predicts the fastest cooling. The first-order theory significantly
under-predicts the work done during the expansion relative to the 
M\"uller-Israel-Stewart and Euler predictions.
Thus the temperature decreases more slowly with the inclusion of
dissipative effects. This would lead to greater
yields of photons and dileptons, and also the transverse energy/momentum would be reduced
as the collective velocities are dissipated into heat. The system takes longer
to cool down. This will delay freeze-out. Also entropy, $s=4\,a\,T^3$, is enhanced.
This is important because entropy production can be related to final
multiplicity. With respect to entropy production due to particle production, see
\cite{azwi:elze}.
Given some initial conditions we want to investigate the importance of second
order theories as compared to first order theories and perfect fluids.
Let us now analyze the differences between the second-order and first-order
theories.
The first thing we notice is that the Eckart-Landau theory predicts
that at early times the temperature will first rise before
falling off. This is more pronounced when we have small initial times. 
 Naively one would
expect that the system would cool monotonically as it expands even in the case of dissipation 
where energy-momentum is conserved. On the other hand, it is
seen that for large initial times and high temperatures the three theories
have a similar time evolution. As can be seen from Fig. 
\ref{azwi:Fig4}, all three cases start at the same point and then fall off 
with time.
The difference stems from the fact that in the second-order theory the
transport equations of the dissipative fluxes describe the evolution of these
fluxes from an arbitrary initial state to an equilibrium state. The
first-order theory, though, is just related to the thermodynamic forces which, if
switched off, do not demonstrate relaxation. Hence they are sometimes
referred to as {\em quasi-stationary} theories.
As can be seen from Fig. \ref{azwi:Fig4}, it is before the establishment of an
 equilibrium state that the two theories differ significantly. 
In ultra-relativistic heavy ion collisions, where the fluid evolution 
occurs in very short times, the second-order theories should be used to analyze 
collision dynamics. 
In doing so a full analysis requires that all the dynamical equations
with more realistic equations of state and transport coefficients be considered

I will conclude by pointing out some of the advantages and
challenges of the second-order theories. Second-order theories, being hyperbolic
in structure, lead to well-posed initial-value (Cauchy) problems. They also lead 
to causal propagation. Unlike the first-order
theories, second-order theories have 
relaxation terms which permit us to study the evolution of the dissipative
fluxes. The challenge we face is the increase in the space of thermodynamic
variables. We now have, in addition to the transport coefficients, 
new coefficients in the problem. These are the relaxation coefficients
$\beta_i$ and the coupling coefficients $\alpha_i$. These new
coefficients depend on the primary thermodynamic variables, such as 
$n,\varepsilon$ and $p$ and therefore are determined by the equation of state.
Like viscosity and thermal conductivity, which are
required to be positive by the second law of thermodynamics, these new
coefficients are 
constrained by hyperbolicity requirements. In principle , in order to solve the
second-order relativistic dissipative fluid dynamic problem, one still needs
the equation of state,
initial conditions and the transport coefficients.

To probe non-equilibrium properties of matter produced in heavy ion
collisions we need a non-equilibrium fluid dynamics model to analyze 
observables.
The relativistic fluid dynamics modeling of 
heavy ion collisions will have to include dissipation and thermal conduction.
One will then have to use the hyperbolic theories of relativistic dissipative
fluids because of their universality. Hyperbolic theories might prove to be
convenient in constructing hydro-molecular dynamic schemes \cite{azwi:bass} in which a
phenomenological fluid dynamics model is coupled to a microscopic kinetic model 
such that microscopic kinematic quantities may be obtained.
Dissipation mechanism might be important if we deal with the event-by-event
based hydrodynamics \cite{azwi:aguiar}.

\acknowledgements
I would like to thank J.I. Kapusta for helpful discussions and preparation of
this work. I would also like to thank D.H. Rischke, L. McLerran, A. Dumitru,
P.J. Ellis for valuable comments.
This work was supported by the US Department of Energy grant
DE-FG02-87ER40382.
\begin{figure}[htbp]
\psfig{file=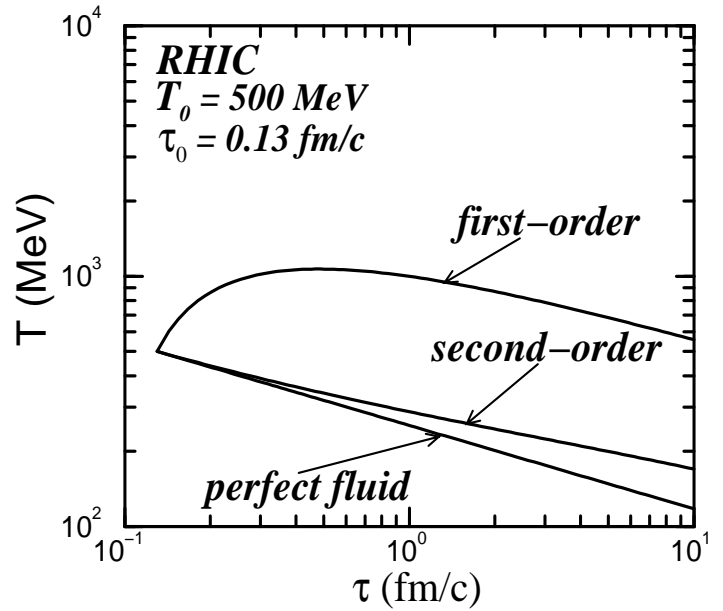,width=4.0in,height=3.5in}
\caption{\small Proper time evolution of temperature for a RHIC
scenario: $\tau_0$ = 0.13 fm/c and $T_0$ = 500 MeV .} \label{azwi:Fig1}
\end{figure}
\begin{figure}[htbp]
\psfig{file=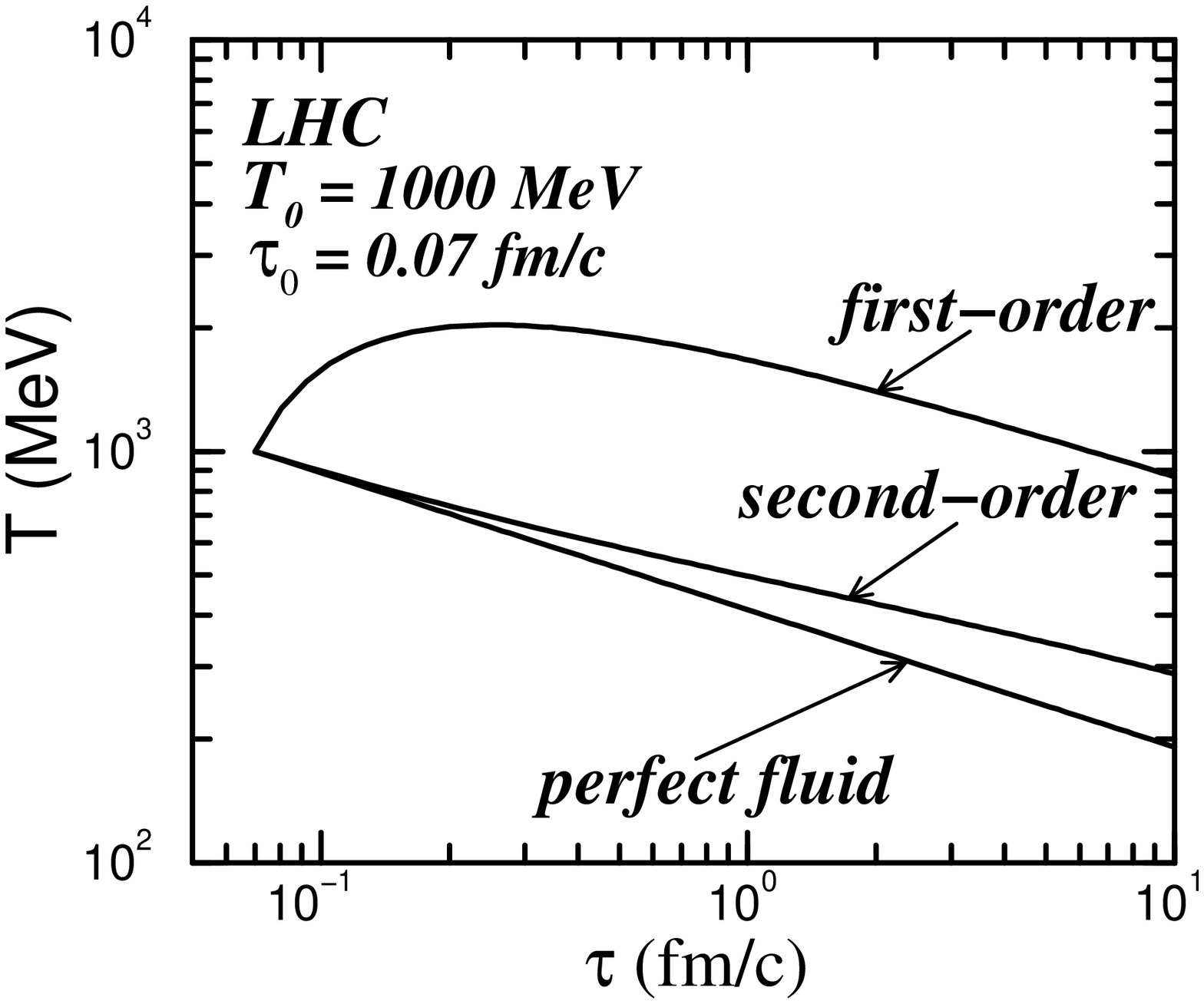,width=4.0in,height=3.5in}
\caption{\small Proper time evolution of temperature for a LHC
scenario: $\tau_0$ = 0.07 fm/c and $T_0$ = 1000 MeV .} \label{azwi:Fig2}
\end{figure}
\begin{figure}[htbp]
\psfig{file=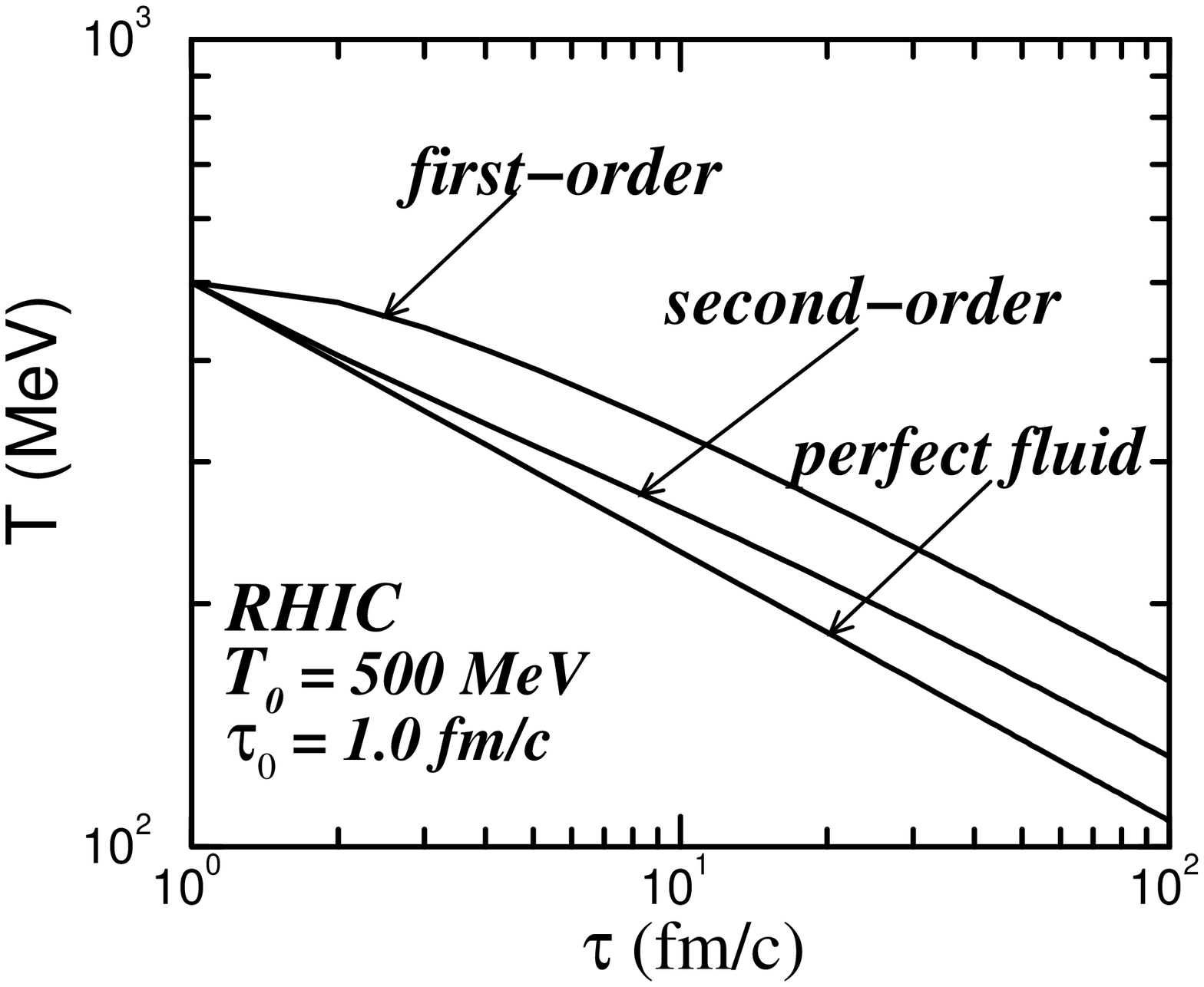,width=4.0in,height=3.5in}
\caption{\small Proper time evolution of temperature for a 
RHIC scenario: $\tau_0$ = 1.0
fm/c and $T_0$ = 500 MeV.} \label{azwi:Fig3}
\end{figure}
\begin{figure}[htbp]
\psfig{file=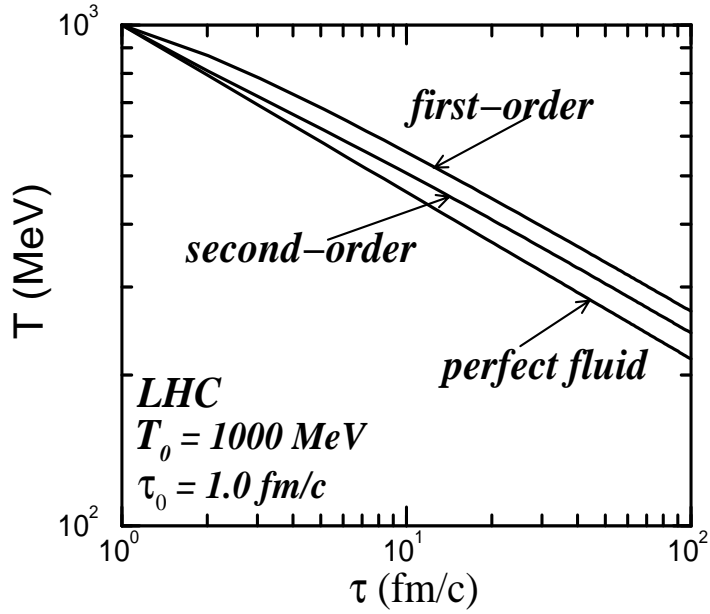,width=4.0in,height=3.5in}
\caption{\small Proper time evolution of temperature for a 
LHC scenario: $\tau_0$ = 1.0 fm/c and $T_0$ = 1000 MeV.} \label{azwi:Fig4}
\end{figure} 

\end{twocolumn}


\begin{references}

\bibitem[*]{byline}  Electronic address: amuronga@physics.spa.umn.edu

%
%
\bibitem{azwi:stocker1986} H. St\"ocker and W. Greiner, Phys. Rep. {\bf 137},
227 (1986); R. B. Clare and D. Strottman, Phys. Rep. {\bf 141}, 177 (1986).
\bibitem{azwi:kajantie1986} K. Kajantie, M. Kataja, L. McLerran and P.V.
Russskanen, Phys. Rev. D {\bf 34}, 2746 (1986);
E.F. Staubo, A.K. Holme, L.P. Csernai, M. Gong and D. Strottman, Phys. Lett. {\bf B229}, 351 (1989);
I.N. Mishustin, V.N. Russkikh and L.M. Satarov, Nucl. Phys. {\bf A494}, 595(1989);
U. Katscher, D.H. Rischke, J.A. Maruhn, W. Greiner, I.N Mishustin and L.M. Satarov,
Z. Phys. {\bf A346}, 209 (1993). 
%
\bibitem{azwi:kapusta} J.I. Kapusta, Phys. Rev. C {\bf 24}, 2545 (1981).
%

%
\bibitem{azwi:eckart1940} C. Eckart, Phys. Rev. {\bf 58}, 919 (1940).
\bibitem{azwi:landau1959} L. D. Landau and E. M. Lifshitz, {\it Fluid
Mechanics}, (Addison-Wesley, Reading, Massachusetts, 1959).
\bibitem{azwi:naviergang} K. Kajantie, Nucl. Phys. {\bf A418}, 41c (1984); 
G. Baym, Nucl. Phys. {\bf A418} 525c (1984); A. Hosoya and K. Kajantie, Nucl. Phys. {\bf
B250}, 666 (1985); P. Danielewicz and M. Gyulassy, Phys. Rev.
D {\bf 31}, 53 (1985); 
L.\ Mornas and U.\ Ornik, Nucl.\ Phys.\ {\bf A587}, 828 (1995);
H. Kouno, M. Maruyama, F. Takagi and K. Saigo, Phys.
Rev. D {\bf 41}, 2903 (1990); R. Hakim and L. Mornas, Phys. Rev. C {\bf 47}, 
2846 (1993).
%
\bibitem{azwi:weinberg} S. Weinberg, {\it Gravitation and Cosmology: Principles and
Applications of the General Theory of Relativity} (J. Willey and Sons, New York, 1972).
\bibitem{azwi:laszlobook}
L.P.\ Csernai, 
{\it Introduction to Relativistic Heavy Ion Collisions}
(J. Wiley and Sons, New York, 1994).
\bibitem{azwi:degroot} S.R. de Groot, H.A. van Leeuven and C.G. van Weert,
{\it Relativistic Kinetic Theory} (North-Holland, Amsterdam, 1980).
%
\bibitem{azwi:rischke} D.H. Rischke, in {\it Hadrons in Dense Matter and
Hadrosynthesis}, J. Cleymans, H.B. Geyer and F.G. Scholtz (Eds.), (Lecture Notes
in Physics, Vol. {\bf 516}, Springer, 1999).
\bibitem{azwi:grad1949} H. Grad, Commun. Pure Appl. Math. {\bf 2}, 331 (1949).

\bibitem{azwi:muller1967} I. M\"uller,  Z. Phys. {\bf 198}, 329 (1967).
\bibitem{azwi:israelstewart}
W.\ Israel, Ann.\ Phys.\ (N.Y.) {\bf 100}, 310 (1976);
J.M.\ Stewart, Proc.\ Roy.\ Soc. {\bf A357}, 59 (1977);
W.\ Israel and J.M.\ Stewart, Ann.\ Phys.\ (N.Y.) {\bf 118},( 341 (1979).
%
\bibitem{azwi:chapman} S. Chapman and T.G. Cowling, {\it The Mathematical Theory of
Non-Uniform Gases}, Third Edition (Cambridge University Press, Cambridge, 1970).
%
\bibitem{azwi:project} Detailed calculations  to be published in PhD thesis.
%
\bibitem{azwi:bjorken} J.D. Bjorken, Phys. Rev. D {\bf 27}, 140 (1983).
%
\bibitem{azwi:hiscock}
W.A.\ Hiscock and L.\ Lindblom, Ann.\ Phys.\ (N.Y.) {\bf 151},  466 (1983);
Phys.\ Rev.\ D {\bf 31}, 725 (1985);
ibid. {\bf 35}, 3723 (1987).
\bibitem{azwi:prakash}
M.\ Prakash, M.\ Prakash, R.\ Venugopalan and G.\ Welke,
Phys.\ Rep.\ {\bf 227}, 321 (1993).
\bibitem{azwi:baym}  G. Baym, H. Monien, C.J. Pethick and D.G. Ravenhall, Phys.
Rev. Lett. {\bf 64}, 1867 (1990).
\bibitem{azwi:kms}  J. Kapusta, L. McLerran and D. K. Srivastava, Phys. Lett.
{\bf B283}, 145 (1992).
%
\bibitem{azwi:elze} H.-Th. Elze, J. Rafelski and L. Turko, Phys. Lett. {\bf
B506}, 123 (2001).
%
\bibitem{azwi:bass}  S.A. Bass, and A. Dumitru, Phys. Rev. C {\bf 61},
064909, 2000.
%
\bibitem{azwi:aguiar} C.E. Aguiar, T. Kodama, T. Osada and Y. Hama, J. Phys. {\bf G27}, 75
(2001).
%
%
%
%
%
%
%
%
%
%
%
%
%
%
\end{references}
\end{document}